\documentstyle[preprint,aps]{revtex}

\newcommand\eqn[1]{\begin{equation}#1\end{equation}}

\title{
Analysis of Molecular Rotational Spectra
}

\author{
K. Hara and G. A. Lalazissis \\
Physik-Department, Technische Universit\"{a}t M\"{u}nchen \\
D-85747 Garching bei M\"{u}nchen, Germany \\
}

\begin{document}
\maketitle

\begin{abstract}
Rotational spectra of diatomic molecules measured in the high-precision
experiments are analyzed. Such a spectrum is usually fitted by an 8th
order polynomial in spin. In fact, from the theoretical point of view,
the rotational spectrum is expected to be a smooth function of spin.
Thus, fitting the rotational spectrum in terms of a low order polynomial
in spin seems to be justified. However, this is not what we found in the
present work. We show that the measured spectrum is not necessarily a
smooth function of the spin, which is quite surprising. Whether or not
it indicates the presence of some hidden physics is an open question.
\end{abstract}

\narrowtext
\newpage
\section{Introduction}
Characteristic quantities of a diatomic molecule are the energy
associated with the motion of electrons and the one associated with
the motion of two nuclei (ionic cores). The latter motion is essentially
that of a dumbbell shaped system (vibration against one another and
rotation as a whole). The order of magnitude of the electronic vs.
vibrational excitation energy ($E_{ele}/E_{vib}$) is typically 100 and
that of the vibrational vs. rotational one ($E_{vib}/E_{rot}$) another
100. It means that $E_{viv}~(E_{rot})$ corresponds to transitions in the
near (far) infrared. Thus, the electronic motion is very fast compared
with that of nuclei, so that its wavefunction may be constructed from
the instantaneous constellation of two nuclei assuming as if nuclei were
frozen and thus depends parametrically on the positions of two nuclei
(Born-Oppenheimer approximation). This is the standard way of decoupling
the electronic and nuclear degrees of freedom from one another. The
treatment of the nuclear part then becomes very simple.

The nuclear radial Schr\"{o}dinger equation (with $\hbar=1$) takes the
form
\eqn{
{1 \over {2M}}{d^2 \psi_{VI}(r) \over {dr^2}} + \left[E_V(I)-U(r)
-{I(I+1) \over {2M r^2}}\right]\psi_{VI}(r)=0
\label{1.1}}
where $r$ is the distance between two nuclei and $M$ the reduced mass.
The potential $U(r)$ between two nuclei is a complicated object that
involves an integration over the electronic wavefunction, which we do
not go into here. In practice, it is phenomenologically replaced by a
Morse potential. The quantum number $I$ is the rotational spin which
characterizes the angular wavefunction $Y_{IM}(\theta,\phi)$ and $V$ the
so-called vibrational quantum number which characterizes the radial
wavefunction. We note that, because of the mathematical property of this
equation, {\em the eigenvalue $E_V(I)$ has to be an analytic function of
the parameter (spin) $I$}. Since the effective potential
\eqn{
V(r)=U(r)+{I(I+1) \over {2Mr^2}}
\label{1.2}}
has a prominent minimum at $r=r_0$, where $r_0$ is of the order of the
molecular size, one usually expands it around $r=r_0$ ($\left[{dV(r)
\over {dr}}\right]_{r=r_0}=0$) and expresses $V(r)$ in a power series of
$r-r_0$. Consequently, the radial motion can be treated as a vibration
having a perturbing unharmonicity. The dependence of $E_V(I)$ on $I$
becomes then a power series of $I(I+1)$ and one usually stops the series
at the 8th order in $I$:
\eqn{
E_V(I)=T_V+B_VI(I+1)-D_V[I(I+1)]^2+H_V[I(I+1)]^3+L_V[I(I+1)]^4
\label{1.3}}

In recent years, very accurate data of rotational spectra in diatomic
molecular bands became available thanks to modern experimental
techniques [1--8]. What is measured in such an experiment is the
so-called R- and P-process \cite{hb}
\eqn{
R(I) = E_H(I+1)-E_L(I), ~~~ P(I) = E_H(I-1)-E_L(I).
\label{1.4}}
These quantities are the `inter-band' $\Delta I=1$ (dipole) transition
energies between a higher ($V=H$) and a lower ($V=L$) vibrational band
and are usually measured in the unit of wave number ($cm^{-1}$).
The coefficients in the formula (\ref{1.3}) have been fitted to such
experimental data [1--8]. However, one can take a different approach
\cite{george}.

By inverting the relation (\ref{1.4}), one obtains the `intra-band'
$\Delta I=2$ (quadrupole) transition energy $\Delta E_V(I) \equiv
E_V(I)-E_V(I-2)$ as
\eqn{
\Delta E_H(I) = R(I-1)-P(I-1), ~~~ \Delta E_L(I) = R(I-2)-P(I)
\label{1.5}}
for the higher and lower band, respectively. It should be remarked that,
from the set of data of R- and P-process, one obtains four separate sets
of data $\Delta E_V(I)$ for $V=H$ and $V=L$ bands with even as well as
odd spin sequences. The intra-band transition energy should be a smooth
function of spin $I$ according to the theoretical consideration made in
the beginning. In the present work, we want to examine this statement.
We will in fact show that the measured intra-band transition energies
are not necessarily smooth functions of spin. Let us first devise a
tool which suits the analysis of data.

\section{The method of analysis}
To study the behavior of $\Delta I=2$ transition energy $\Delta E(I)=
E(I)-E(I-2)$ for each band ($V=H$ or $L$) and each spin sequence
($I=$even or odd), some kind of data manipulation is necessary. In
fact, if one plots the quantity $\Delta E(I)$ directly, it is hardly
possible to see any fine structure since $\Delta E(I)$ is a globally
increasing function of $I$ and extends over a wide range of values.
Thus, the basic idea is to look at the deviation of the transition
energy $\Delta E(I)$ from its smoothly increasing part.

To this purpose, we subtract a polynomial of order $N$ in $I$ from
$\Delta E(I)$ and define what we call the $N$th order one-point formula
\cite{hl}
\eqn{
\Delta^{(1)}_{N} E(I) \equiv \Delta E(I) - Q_N(I),
~~~Q_N(I) = \sum_{m=0}^N q_m I^m
\label{2.1}}
where the coefficients $q_m$ are determined by minimizing the quantity
\eqn{
\chi(q_0,\cdots,q_N) \equiv \left[\Delta^{(1)}_{N} E(I)\right]^2
\label{2.2}}
with respect to $q_m$ (${\partial \chi \over {\partial q_m}}=0$). This
leads to a set of $N+1$ equations ($m=0,1,\cdots,N$)
\eqn{
\sum_{n=0}^N S_{m n} q_n = T_m,
~~~S_{m n} \equiv \sum_I^{\Delta I=2} I^m I^n,
~~~T_m \equiv \sum_I^{\Delta I=2} I^m \Delta E(I).
\label{2.3}}
We note that the smooth part $Q_N(I)$ which we subtract from $\Delta
E(I)$ is nothing other than a polynomial of order $N$ determined by the
$\chi$-square fit to $\Delta E(I)$. However, in practice, this formula
cannot be used in the present form particularly when the order of the
polynomial $N$ is larger than 3 since the equation (\ref{2.3}) is highly
ill-conditioned. Thus, in what follows, we want to transform it into
another form.

First, let us note that the replacement $I \rightarrow I-I_0$ does not
change the fitting procedure since the shape of the polynomial $Q_N(I)$
as a function of $I$ is unchanged. Thus, the origin of the spin values
can be shifted freely. Secondly, the spin variable can be scaled too ($I
\rightarrow aI$) since the order of the polynomial remains the same.
These properties can be used to rewrite the polynomial in a different
form.

Shifting and scaling the spin values can be achieved most generally by a
linear mapping $I=ax+b$. The increment of $x$ is thus $\Delta x={\Delta
I \over a}$ ($\Delta I = 2$). We will choose $a={I_{max}-I_{min} \over
2}$ and $b={I_{max}+I_{min} \over 2}$ so that the range of $x$ becomes
[-1,+1], where $x=-1~(+1)$ corresponds to $I=I_{min}~(I_{max})$. The
polynomial in question may thus be written in the form
\eqn{
Q_N(I)=\sum_{m=0}^N p_m P_m(x).
\label{2.4}}
Here, we use the Legendre polynomial $P_m(x)$ instead of $x^m$. The
reason will be explained below. The resulting set of equations is
similar to (\ref{2.3}) but $q_n$ is replaced by $p_n$ and $I^n$ ($I^m$)
by $P_n(x)$ ($P_m(x)$). This representation has an advantage that there
holds the relation
\eqn{
S_{m n} = \sum_{x=-1}^{+1} P_m(x) P_n(x) = 0 ~~~ {\rm if} ~m+n=odd.
\label{2.5}}
It means that the whole set of equations splits into two independent
sets of equations of smaller dimensions, one with $m,n=$even and the
other with $m,n=$odd:
\eqn{
\sum_{n={\rm even~or~odd}}^N S_{m n} p_n = T_m,
~~~S_{m n}\equiv\sum_{x=-1}^{+1} P_m(x) P_n(x),
~~~T_m\equiv\sum_{x=-1}^{+1} P_m(x) \Delta E(I).
\label{2.6}}
This set of equations determines the coefficients $p_m$ and accordingly
the polynomial (\ref{2.4}) which represents the smooth part of $\Delta E(I)$.

It remains to be remarked that the original set of equations (\ref{2.3})
is highly ill-conditioned. It is indeed so ill-conditioned that even the
double precision algorithm is not free from the numerical instability
caused by large losses of accuracy if $N$ is greater than 3. This
problem can be avoided by shifting and scaling the spin values as
presented above. In fact, the situation improves slightly if one uses
the power series $x^m$ thanks to a property analogous to (\ref{2.5}).
Nevertheless, this does not fully resolve the numerical instability. The
reason lies basically in the fact that the zeros of $x^m$ are multiple
and are all concentrated at $x=0$. In contrast, all zeros of $P_m(x)$
are simple and never coincide with one another for different $m$'s. The
use of the Legendre polynomial $P_m(x)$ instead of the power series
$x^m$ is thus essential for the numerical reliability (both accuracy and
stability). A similar caution may be necessary when fitting the formula
(\ref{1.3}) to experimental data.

In the next section, we will present $\Delta^{(1)}_{N} E(I)$ for a
number of experimental data using a 7th order polynomial. In fact, $N=7$
is quite appropriate for the analysis of molecular rotational spectra
since the rotational spectrum for a given vibrational band is fitted
(globally) by a formula containing up to the 8th order in spin $I$ (see
eq.(\ref{1.3})) so that the transition energy $\Delta E(I)$ is a 7th
order polynomial.

We will also define the `filtered' one-point formula by setting the
quantity $\Delta^{(1)}_{N} E(I)$ to zero if its absolute value is
smaller than or equal to the corresponding error bar. This formula is
quite useful in practice. By construction, it shows whether the
deviation of the transition energy $\Delta E(I)$ from its smooth part is
physically significant or not.

\section{Analysis of experimental data}
As mentioned before, we obtain four sets of transition energies $\Delta
E(I)$ (even and odd spin sequence for higher and lower band) from
the measurement of $R(I)$ and $P(I)$. Thus, four independent sets of
data $\Delta^{(1)}_{N} E(I)$ can be created, which we present as
diagrams.

As an example, we plot in Fig.1a the 7th order one-point formula
$\Delta^{(1)}_{\,7} E(I)$ obtained from the measured $R(I)$ and $P(I)$
of the 2-2 band of the $A^1\Sigma^+-X^1\Sigma^+$ system of the molecule
YN \cite{yn}. On the other hand, Fig.1b shows the corresponding
filtered one-point formula. It is `filtered' in the sense that we take
\eqn{
\Delta^{(1)}_{N} E(I) = 0 ~~~{\rm if}~~~ |\Delta^{(1)}_{N} E(I)|
\le {\rm Error~Bar}
\label{3.1}}
where Error Bar is twice the error bar of the measurement of $R(I)$ and
$P(I)$, see eq.(\ref{1.5}). Eq.(\ref{3.1}) makes easily visible that
some deviations of the transition energy from the smooth part that occur
in Fig.1a are physically insignificant as they are zero within the
Error Bar. Thus, if the rotational spectrum is indeed a smooth function
of spin $I$, which is what we want to confirm, the deviation should be
zero everywhere when filtered. However, Fig.1b clearly shows that there
remain some deviations which are not filtered away to zero ($V=H$ band).
These deviations are beyond the Error Bar and may thus be considered as
significant. This is rather surprising but is not a single case. We
found many such cases, most of them being more complicated than this
example. It should be emphasized that the filtered formula clearly shows
at which spins the anomaly occurs.

We present another example. Figs.2a and 2b show respectively the
one-point formula and the corresponding filtered formula applied to the
1-1 band of the $C^1\Sigma^+-X^1\Sigma^+$ system of the molecule YD
\cite{yd}. One sees a behavior similar to the above example, though it
looks slightly complex. As a matter of fact, this kind of results is
seen also in the 0-0 and 1-0 bands of the $A^6\Sigma^+-X^6\Sigma^+$
system of the CrD molecule \cite{crd}. Actually, there are many more
complex cases in the data we have analyzed, some of which will be
shown later. Thus, it seems that there exists anomaly (irregularity)
in the measured spectrum and the question is where it comes from.

We have examined the relation between a spectrum and the resulting
irregularity which appears in the transition energy. Suppose that there
is a band crossing in a spectrum. It will produce a kink at the crossing
point and thus an irregularity in the transition energy. While it is
unlikely that a band crossing may occur in the molecular cases, it
indeed exists in the nuclear cases \cite{hl}, which leads to the
so-called backbending phenomenon. Based on this observation, we have
studied a model in which a smooth rotational spectrum $E(I)=AI(I+1)$
has a sequence of kinks around a spin $I=I_0$. Since the smooth part of
the transition energy $\Delta E(I)=E(I)-E(I-2)=2A(2I-1)$ is a first
order polynomial in $I$, it will be sufficient to use the 1st order
one-point formula $\Delta^{(1)}_{\,1} E(I)$ in such a demonstrative
example. Some of them are presented in Fig.3.

The three-kink model in Fig.3 is of particular interest to the YN
molecule \cite{yn} as it reproduces the $V=H$ even spin data shown in
Fig.1b while the $V=H$ odd spin data corresponds to a special four-kink
model. The three-kink spectrum in Fig.3 was produced by shifting `down'
the energy from a smooth curve $E(I)=AI(I+1)$ at a spin $I=I_0$ while
the pattern with exactly opposite phase will be obtained by shifting
`up' the energy. In general, a spectrum with a given number of kinks can
produce several different patterns depending on the way how the kinks
are created.

Let us next examine the YD molecule \cite{yd} presented in Fig.2b.
The $V=H$ even spin data can also be understood in terms of a three-kink
model, in which three-fold kinks occur at two different places (at
$I=18$ and $32$ in a spectrum of the form $E(I)=AI(I+1)-B[I(I+1)]^2$).
Fig.4 compares the data and such a `theory'. The experiment is
reproduced quite well. On the other hand, the $V=H$ odd spin data seems
to suggest the presence of different types of kinks that occur at three
different places. In fact, the first one (at the lowest spin region)
shows obviously a two-kink pattern while the last one (at the highest
spin region) a four-kink pattern, cf. Fig.3. Finally, the middle one is
more complicated but one can easily guess that it is a six-kink pattern.

Therefore, whatever the reason may be, it is certain that there exist
various types of kinks that occur locally at different places in these
rotational spectra. In other words, {\em they are for sure not smooth
(or analytic) functions of the spin}.

\section{Conclusion}
We have shown that the molecular rotational spectrum is not necessarily
a smooth function of spin according to the measurements of $R(I)$ and
$P(I)$ as well as the error estimate of the measurements. This is quite
surprising because of our theoretical understanding that the spectrum
has to be an analytic function of spin $I$, which follows from the
equation (\ref{1.1}) obtained by assuming the Born-Oppenheimer
approximation. Since it is rather difficult to believe that this
approximation is violated, the first thought occurred to us was that
there should be some error in the input file to our Fortran code because
even a single error in typing data would create `artificial' kinks. It
may also be worth reconfirming all data at different laboratories with
different experimental setups.

In the previous section, we have shown relatively simple examples but
there are actually many more complex cases. Fig.5 shows the filtered
one-point formula applied to the 0-0 band belonging to the
$B^1\Sigma^+_u-X^1\Sigma^+_g$ system of the $^{63}$Cu$_2$ molecule
\cite{cu}. Nevertheless, it is likely that this result can also be
decomposed and classified into various patterns as we have done in the
previous section. We believe that this way of understanding the
experimental data is suggestive and that intensive (phenomenological)
studies along this line will give a useful clue. The aim of the present
work is to initiate such a study\footnote{The Fortran source code used
in the present work may be distributed upon request. Send an email to
khara@physik.tu-muenchen.de or glalazis@physik.tu-muenchen.de.}. One
day, we may then be able to find the real physical implication lying
behind this `new' phenomenon, although whether or not it contains
something new is an open question.


\newpage
\parindent=2truecm
\centerline{\large FIGURE CAPTIONS}
\begin{description}

\item[Fig. 1] (a) Seventh order one-point formula and (b) the
corresponding filtered formula applied to the YN data \cite{yn}

\item[Fig. 2] (a) Seventh order one-point formula and (b) the
corresponding filtered formula applied to the YD data \cite{yd}

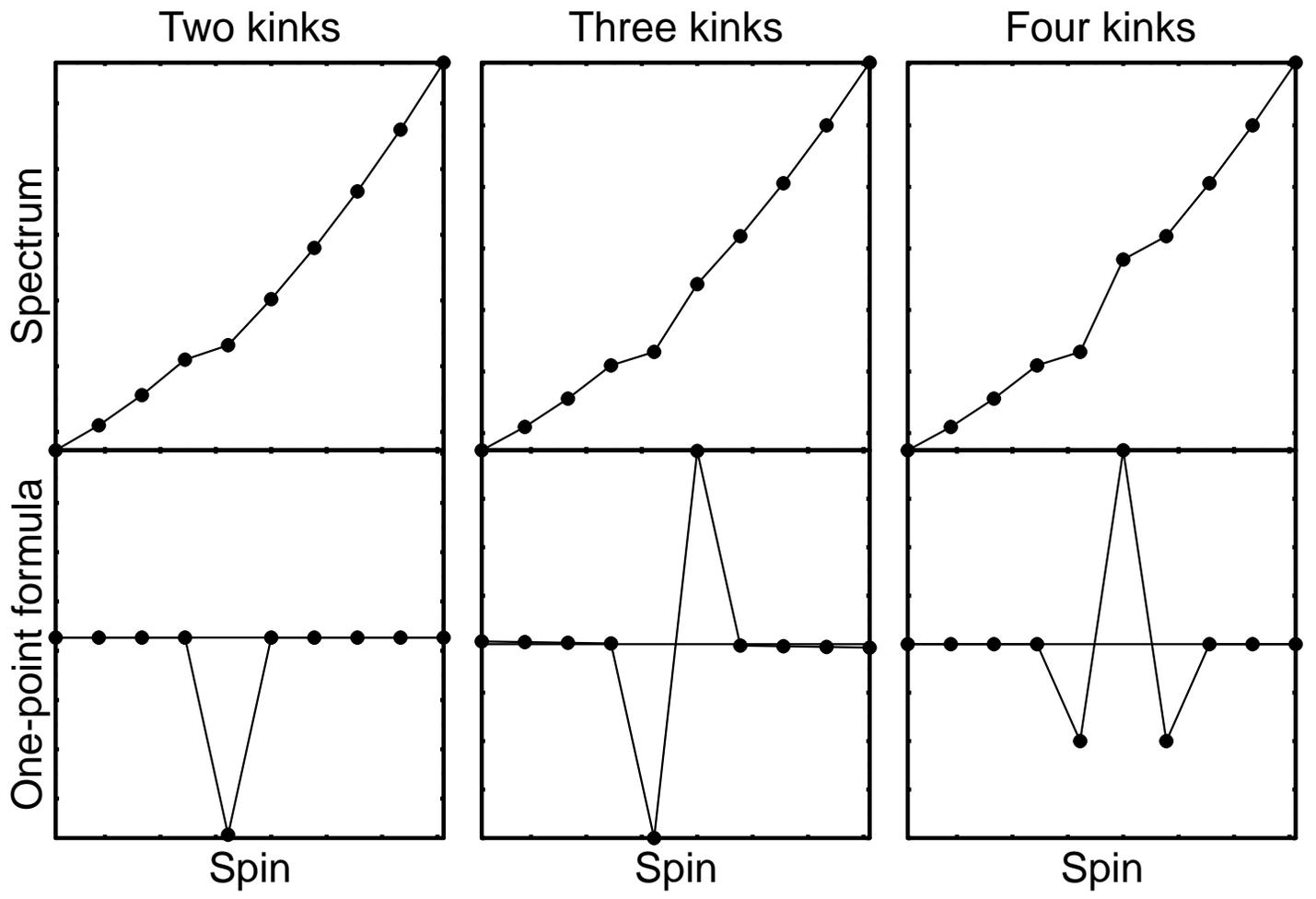
\item[Fig. 3] Model spectra that have two, three and four successive
kinks and the irregularities that occur in the one-point formula

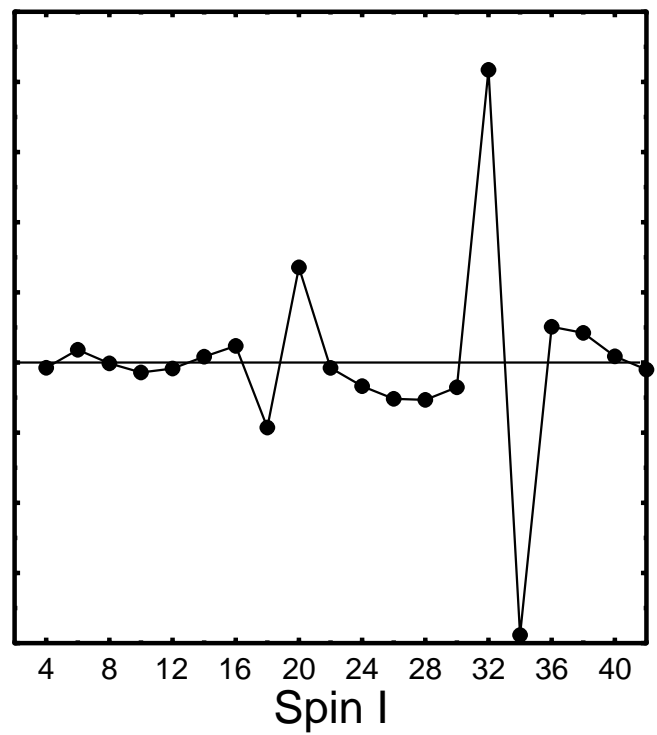
\item[Fig. 4] Comparison between the $V=H$ even spin data of the
YD molecule (cf. Fig. 2b) and a three-kink `theory'

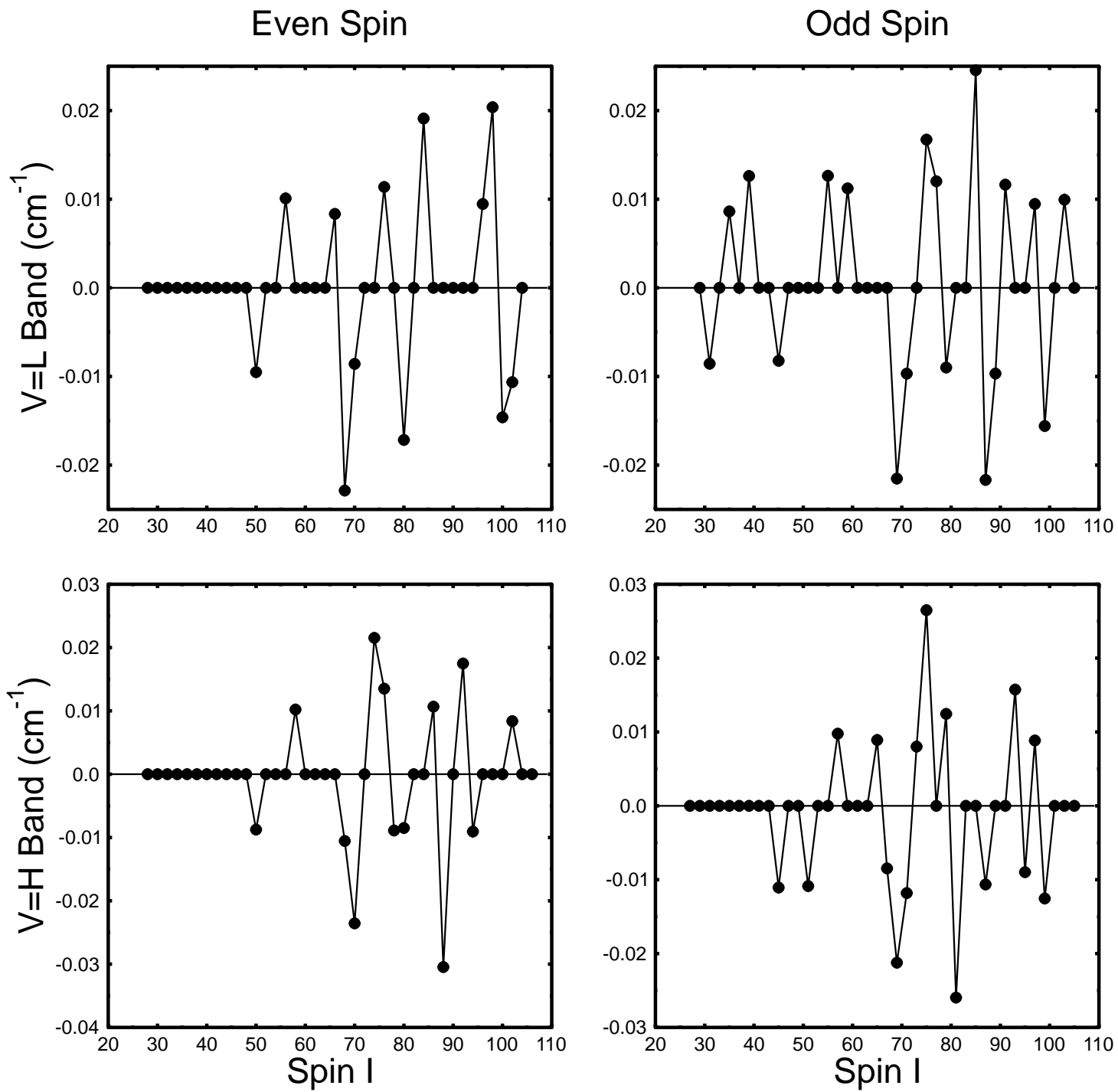
\item[Fig. 5] Seventh order filtered one-point formula applied to the
$^{63}$Cu$_2$ data \cite{cu}

\end{description}

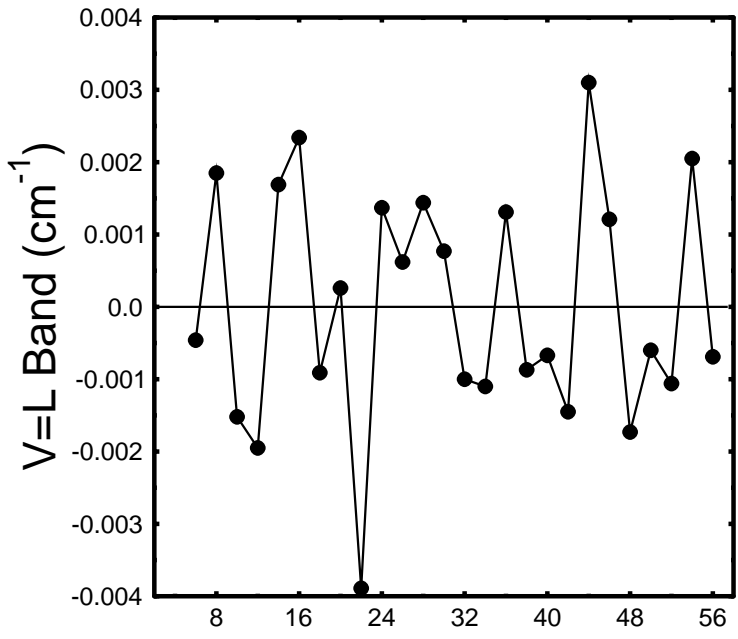
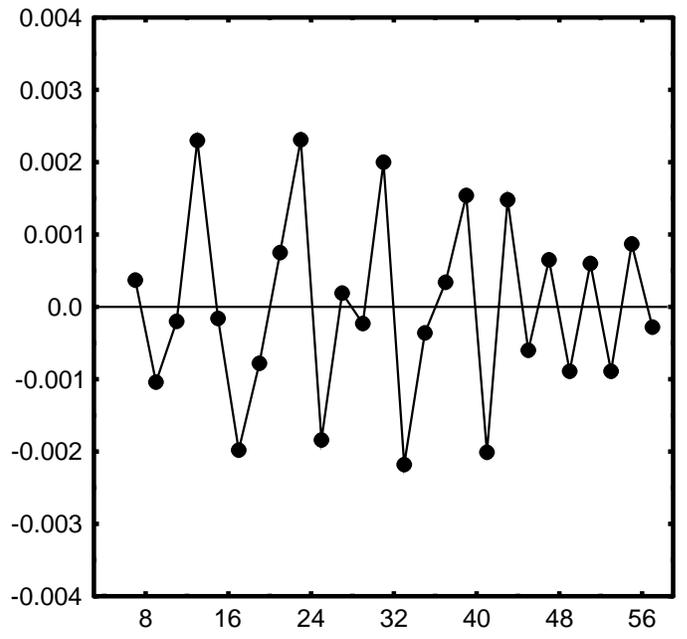
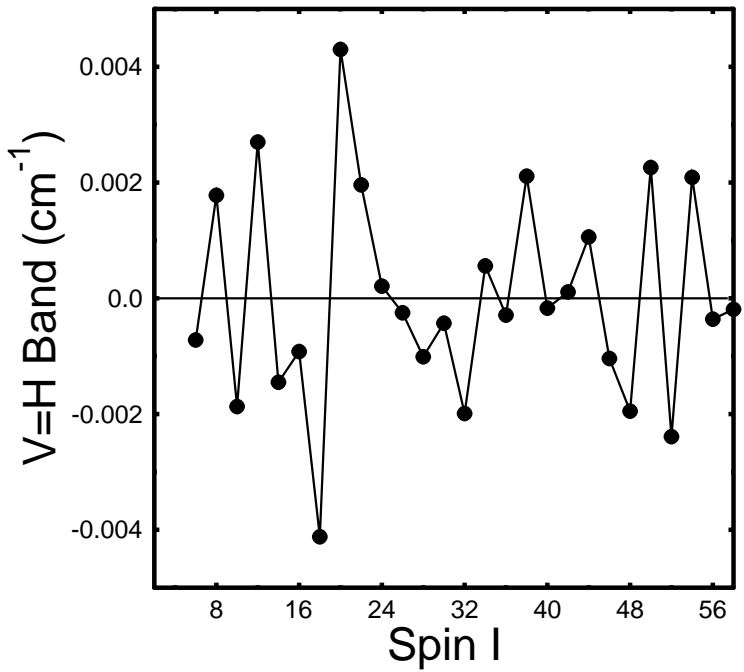
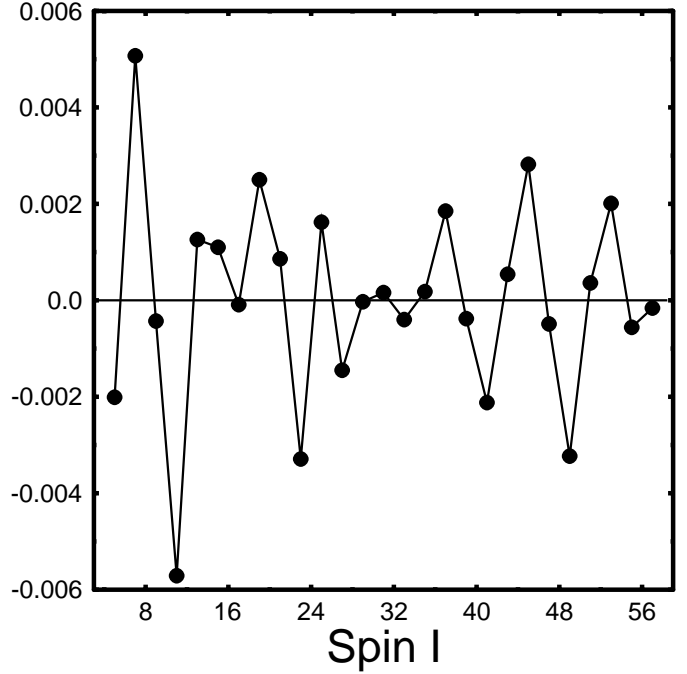
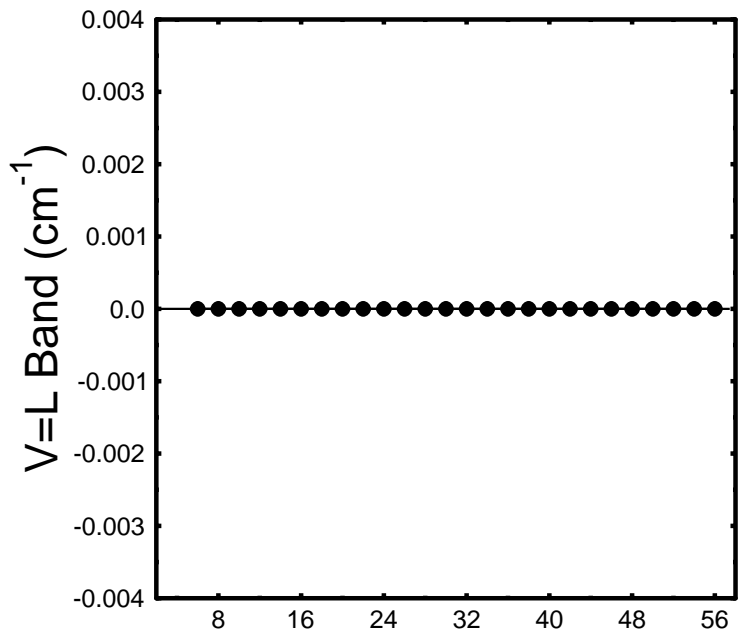
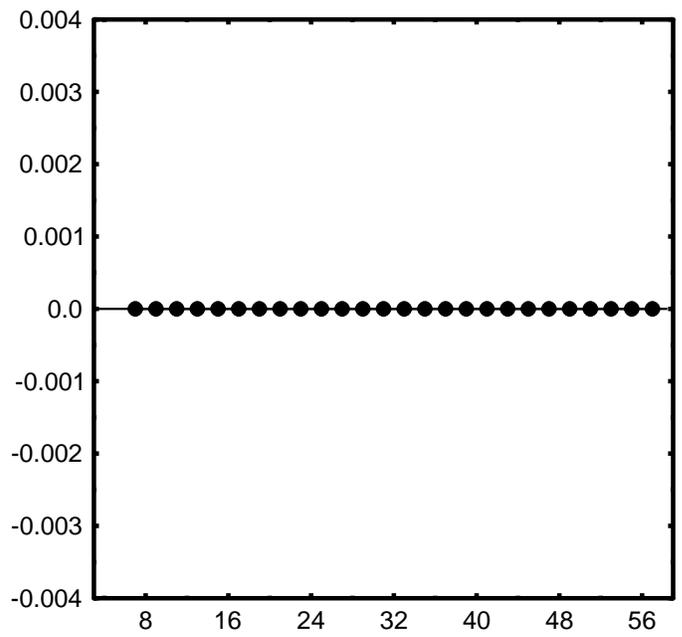
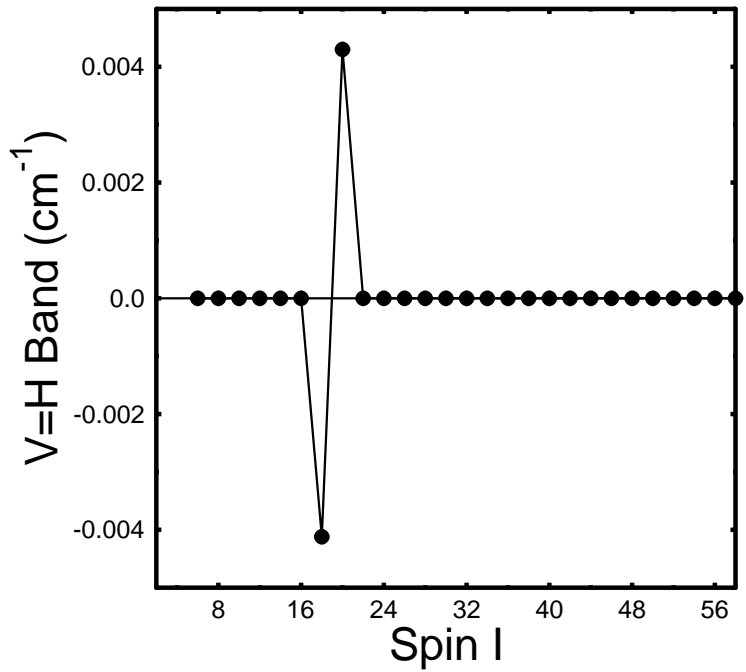
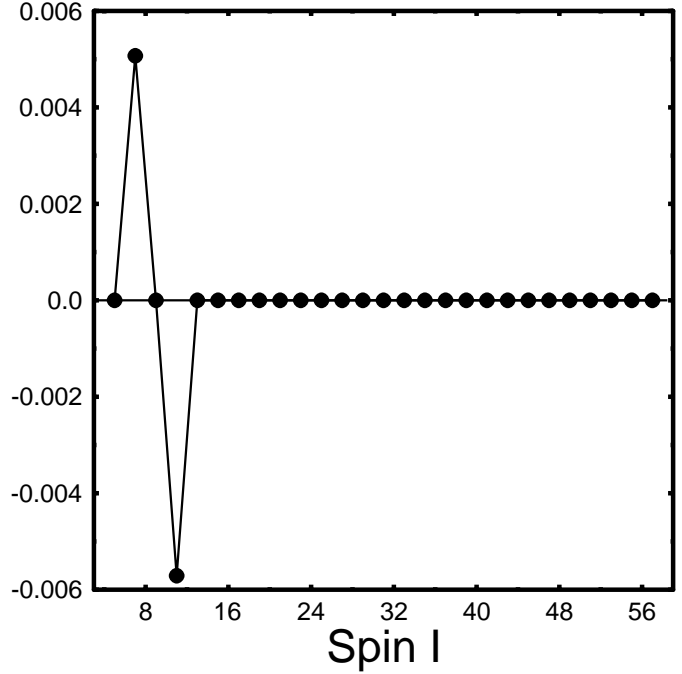
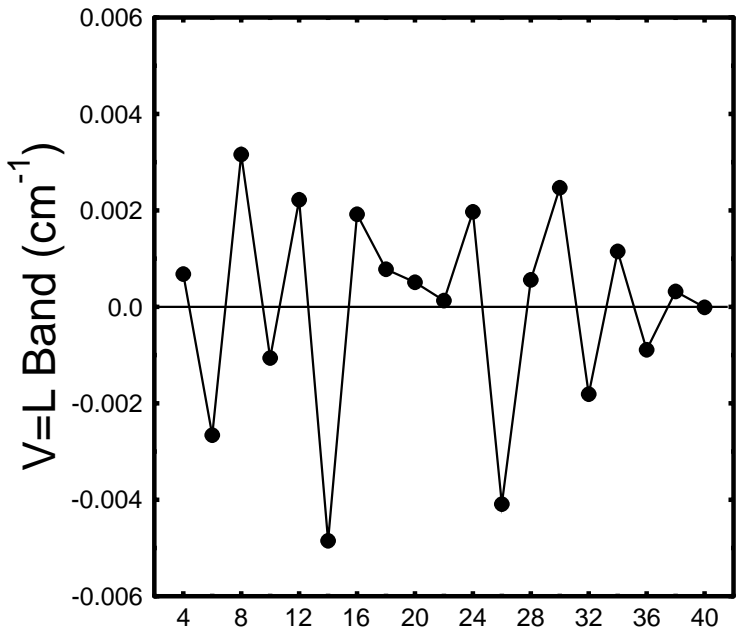
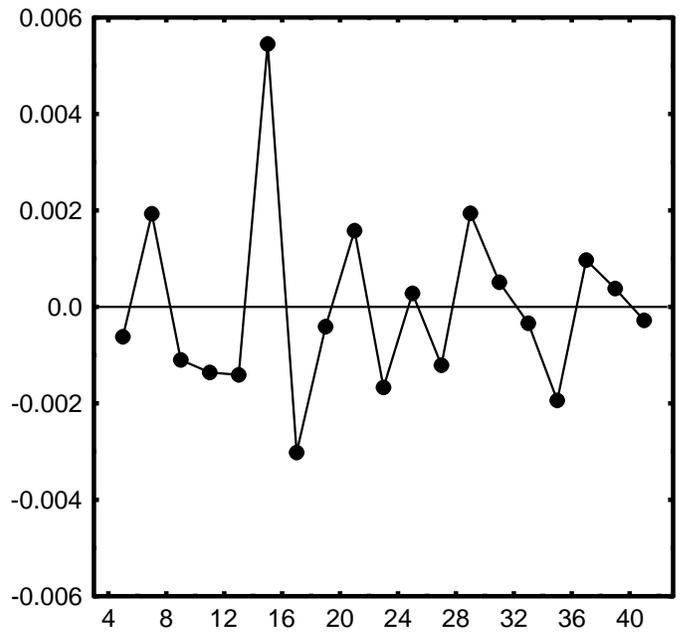
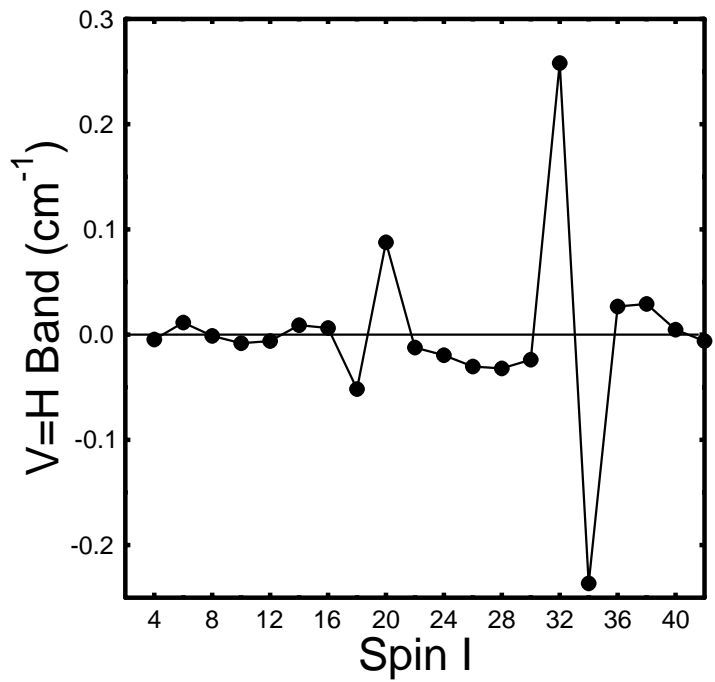
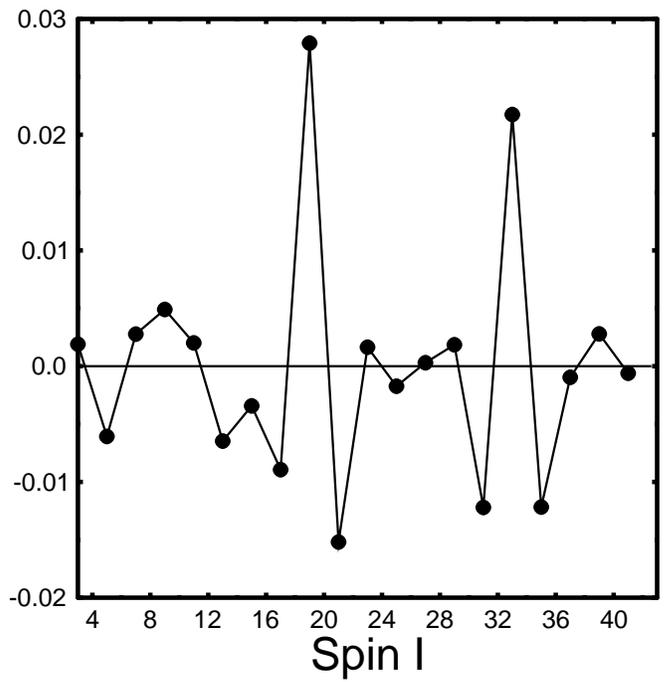
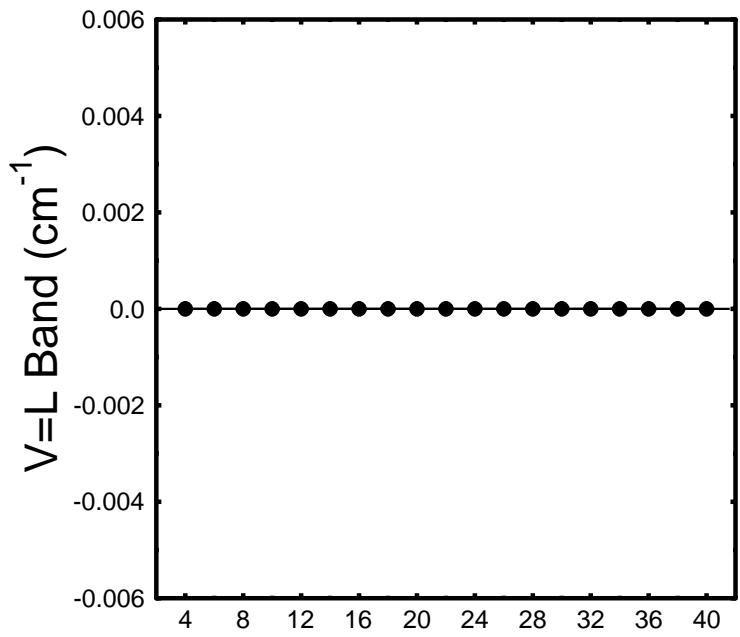
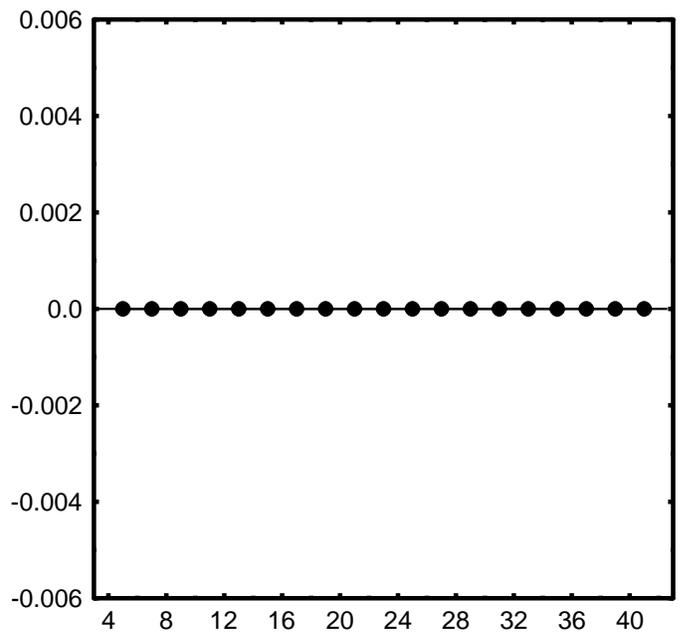
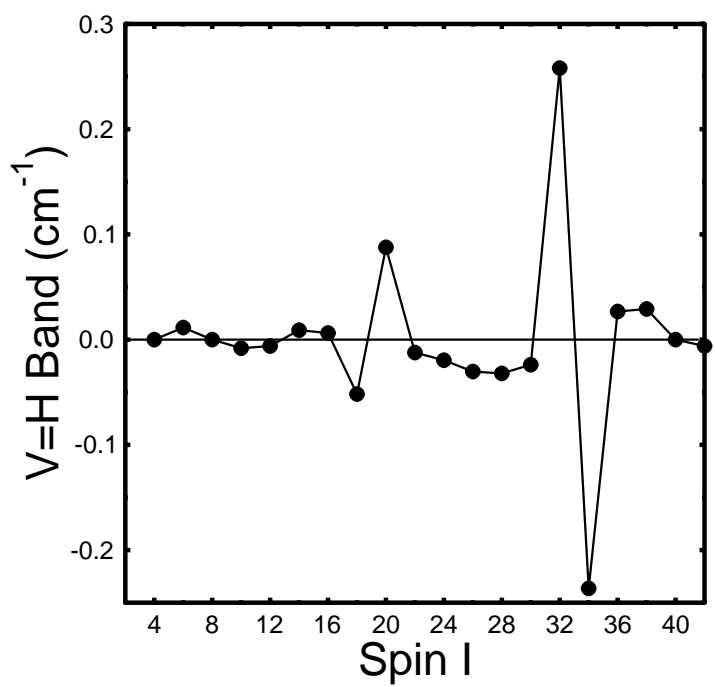
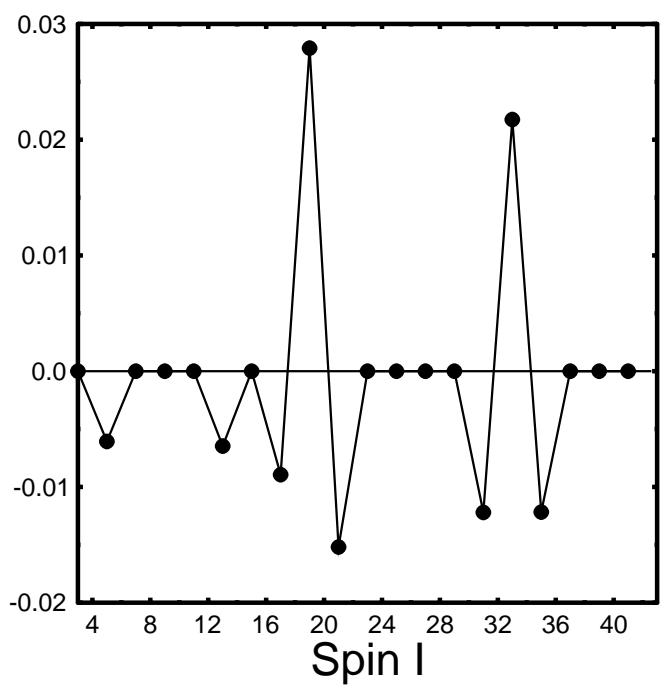
\end{document}